\documentclass[letterpaper,11pt]{article}

\usepackage[T1]{fontenc}
\usepackage[utf8]{inputenc}
\usepackage{newtxtext,newtxmath}
\usepackage{geometry}
\usepackage{setspace}
\usepackage[
  backend=biber,
  style=chem-acs,
  sorting=none,
  articletitle=true
]{biblatex}

\usepackage{graphicx}
\usepackage{float}
\usepackage{amsmath,bm}
\usepackage{booktabs,array,multirow,dcolumn}
\usepackage{authblk}
\usepackage{xurl}
\usepackage[hidelinks]{hyperref}

\newcommand*{\Li}{\ensuremath{\mathrm{Li}^{+}}}
\newcommand*{\PF}{\ensuremath{\mathrm{PF}_{6}^{-}}}
\newcommand*{\ESPmin}{\ensuremath{\mathrm{ESP}_{\min}}}
\newcommand*{\ESPmax}{\ensuremath{\mathrm{ESP}_{\max}}}

\title{Extracting a nitrile-centered, ether-assisted motif hierarchy for lithium-battery electrolyte design from billion-scale molecular space}

\author[1,2,3]{Yifeng Xia}
\author[1,2,3]{Guanghui Wang}
\author[4]{Sining Wang}
\author[4]{Wenting Chen}
\author[5]{Zheng Cheng\textsuperscript{*}}
\author[1,6,2,3]{Jinzhe Zeng\textsuperscript{*}}
\author[1,2,3]{Qiangqiang Gu\textsuperscript{*}}

\affil[1]{School of Artificial Intelligence and Data Science, University of Science and Technology of China, Hefei 230026, China}
\affil[2]{Suzhou Institute for Advanced Research, University of Science and Technology of China, Suzhou 215123, China}
\affil[3]{Suzhou Big Data \& AI Research and Engineering Center, Suzhou 215123, China}
\affil[4]{Contemporary Amperex Technology Co. Limited, Ningde 352100, China}
\affil[5]{AI for Science Institute, Beijing 100080, China}
\affil[6]{State Key Laboratory of Precision and Intelligent Chemistry, University of Science and Technology of China, Hefei 230026, China}

\date{\textsuperscript{*}Corresponding authors: chengz@aisi.ac.cn; jinzhe.zeng@ustc.edu.cn; guqq@ustc.edu.cn}

\begin{document}
\maketitle

\begin{abstract}
Designing electrolyte molecules for lithium batteries requires balancing electronic stability with appropriate \(\mathrm{Li}^+\) solvation, yet the structural basis remains unclear across chemically diverse molecules. High-throughput screening expands the searchable space, but ranked candidates alone do not reveal recurring motifs or their applicability limits. We searched nearly one billion GDB13 structures using electronic--solvation descriptors without explicit functional-group preferences or scaffold constraints. Across descriptor weights, high-ranking populations separated into a nitrile-dominant regime and a coexistence regime containing substantial fractions of both nitrile- and ether-containing molecules. These regimes together define a nitrile-centered, ether-assisted motif hierarchy: nitrile remains favored across broad weight ranges, whereas ether becomes prominent under stronger electrostatic and polarity constraints. Encoding this hierarchy in a generative model expands the candidate space beyond GDB13 and yields high-scoring fluorinated structures without an explicit fluorination reward. Explicit-solvent molecular dynamics simulations show weak, exchangeable coordination of representative candidates without displacing ethylene carbonate from the dominant first solvation shell around \(\mathrm{Li}^+\); effects on ion association and transport depend on molecular structure and concentration. These results establish a quantitative, interpretable and physically bounded motif hierarchy that systematizes established nitrile and ether chemistry for lithium-battery electrolyte design.
\end{abstract}

\section*{Keywords}
lithium battery electrolytes; electrolyte molecule design; electronic--solvation balance; nitrile--ether motif hierarchy; molecule generation; weak coordination; high-throughput molecular screening

\section{Introduction}

Electrolytes in lithium batteries govern \Li transport and shape \Li solvation and cation--anion association\cite{Yao,Chen2020,Xiao2023}, while also participating in interfacial electron transfer and the formation and evolution of the solid- and cathode-electrolyte interphases\cite{Xu2004,Xu2014,Gauthier2015,Borodin2022}. The molecular features required for electronic stability and appropriate solvation are not necessarily aligned.
Nor is stronger solvation invariably beneficial: strong \Li coordination can promote salt dissociation but hinder interfacial desolvation\cite{Ong2015,Fan2024}, whereas weak coordination can ease desolvation at the cost of greater ion pairing and aggregation, thereby reducing the population of mobile charge carriers\cite{Mao2023,Fan2024,Li2025}.
Rational molecular design therefore requires quantitative rules that capture these trade-offs and define the conditions under which they remain valid.

Empirical electrolyte design has long relied on functional groups to control electronic stability and \Li solvation. Fluorination is widely used to tune oxidative stability and interphase formation\cite{Fan2021,Li2023,liu2024,Michan2016}, whereas ether and carbonyl O sites are introduced to modify \Li solvation\cite{Su2019,Xiao2023}. Unsaturated bonds, borate and phosphate units have also been used to regulate redox stability and interphase chemistry\cite{Zhang2006,Fan2021}, while nitrile groups can affect salt dissociation, solvation structure and interphase reactions\cite{Che2025,Yang2025Nitrile}. Most of these structure--function relationships, however, have been established for particular molecules, salt--solvent combinations or performance targets. A functional group that performs well in several formulations may reflect a broader structural principle, or its apparent benefit may depend on the surrounding scaffold and electrolyte composition. It remains unclear which structural motifs are consistently favored when chemically diverse molecules are evaluated under the same electronic and solvation constraints. Resolving this question requires a systematic, functional-group-agnostic search across a chemically diverse molecular space.

Such a search is now feasible because high-throughput computation and machine learning extend systematic analysis to molecular spaces beyond the reach of experimental trial and error. Across large molecular libraries, high-throughput quantum chemistry can evaluate electronic and solvation-related properties\cite{Cheng2015,Gao2023}, while machine learning, active learning and generative models can prioritize or generate candidates for further study\cite{dave2022,ma2025,wang2025,chen2025}. Much of this work has focused on property prediction, candidate ranking and formulation optimization. These outputs identify molecules for further study, but they do not by themselves establish which motifs recur under multiple constraints, why those motifs are selected, or over what range of conditions those patterns persist. Comparing the molecular populations selected under different property priorities can expose recurring motifs that remain obscured in a single ranked list, thereby providing a basis for quantitative design rules and their applicability limits.

In this work, we searched nearly one billion GDB13\cite{blum2009} structures without explicit functional-group preferences or scaffold constraints to identify motifs that recur under competing electronic and solvation requirements. We predicted electronic--solvation descriptors using molecular-property models trained on approximately $10^5$ PubChem\cite{pubchem2021} molecules labeled by density functional theory (DFT). Across different descriptor weights, the high-ranking molecular populations separated into two connected regimes: one dominated by nitrile-containing molecules and another in which nitrile- and ether-containing molecules coexist. These two regimes define a nitrile-centered, ether-assisted motif hierarchy. Structural interpretation indicated that nitrile remains the robust core motif under constraints on electronic stability and local electrostatic potential, whereas ether becomes favored when greater weight is placed on solvation and molecular polarity. Encoding this hierarchy in a generative model expanded the candidate space beyond the enumerated GDB13 library. To test its condensed-phase implications under explicit solvation competition, we performed molecular dynamics (MD) simulations using representative nitrile-, ether- and nitrile--ether-containing additives in an ethylene carbonate (EC)-based electrolyte containing 1.0 M Li\PF, including a fluorinated nitrile candidate generated by the model. The additives exhibited weak, exchangeable coordination through N and O sites without displacing EC from the dominant first-shell environment, thereby defining a condensed-phase boundary for the rule. Together, these analyses translate empirical functional-group knowledge into a quantitative, testable and physically bounded rule for lithium-battery electrolyte design.

\section{Results and Discussion}
We first establish an electronic--solvation descriptor space and validate its scalable prediction across GDB13. We then compare the molecular populations selected under different descriptor weights and determine the structural basis of the resulting nitrile-centered, ether-assisted motif hierarchy. Finally, rule-guided generation tests whether this hierarchy can expand the candidate space, while molecular dynamics simulations delineate its condensed-phase behavior and applicability limits.

\subsection{Electronic--solvation descriptor space and model validation}

\begin{figure}[htbp!]
  \centering
  \includegraphics[width=\textwidth]{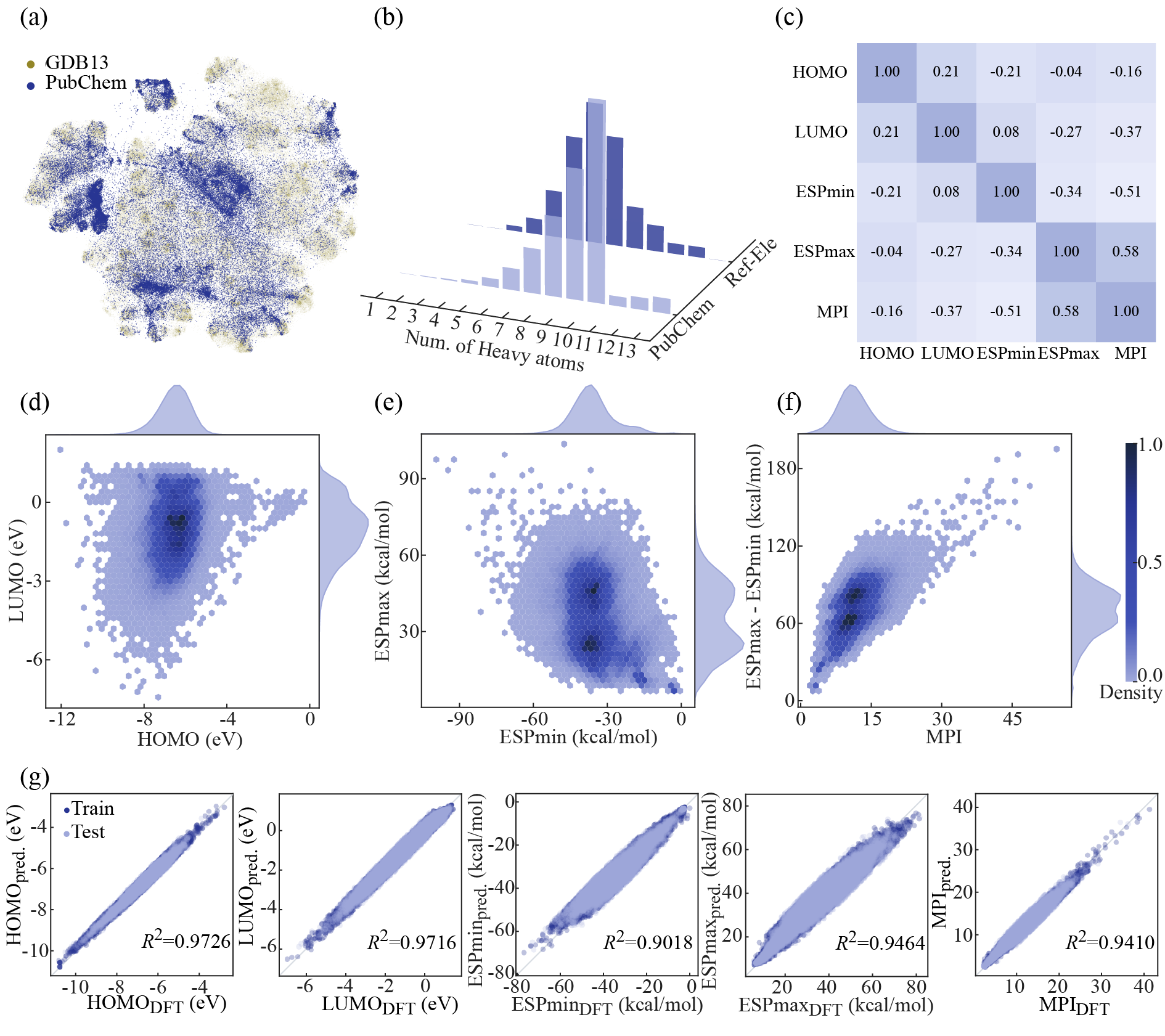}
  \caption{Electronic--solvation descriptor space and surrogate-model validation.
  (a) Two-dimensional t-SNE projection of MACCS fingerprints for sampled GDB13 molecules and PubChem-derived DFT training molecules.
  (b) Heavy-atom-number distributions of the PubChem-derived DFT training set and the reference electrolyte molecule set.
  (c) Pearson correlation matrix of the five DFT-level proxy descriptors: HOMO, LUMO, \ESPmin, \ESPmax, and MPI.
  (d--f) Joint density distributions of the PubChem-derived DFT dataset for HOMO versus LUMO, \ESPmin{} versus \ESPmax{}, and MPI versus the electrostatic-potential range (\ESPmax{} minus \ESPmin{}), respectively.
  (g) Parity plots between DFT-calculated and Uni-Mol-predicted values for the five descriptors in the PubChem training and held-out test sets; independent validation on GDB13 molecules is shown in Fig.~S3.}
  \label{fig:dataset}
\end{figure}

GDB13 comprises nearly one billion enumerated organic molecules containing no more than 13 heavy atoms, providing a large statistical molecular space in which to identify structural regularities beyond known electrolyte compounds\cite{blum2009}. Direct evaluation of condensed-phase solvation structures, interfacial reaction pathways or free-energy changes is infeasible at this scale. We therefore considered five isolated-molecule properties defined at the DFT level as tractable screening coordinates. The highest occupied and lowest unoccupied molecular orbital (HOMO and LUMO) energies serve as proxies for intrinsic oxidation and reduction tendencies\cite{Cheng2015,Pande2019,Gao2023,Peljo2018}. The extrema of the molecular surface electrostatic potential, \ESPmin and \ESPmax, characterize locally electron-rich and electron-deficient surface regions, whereas the molecular polarity index (MPI) summarizes the overall heterogeneity of the surface potential\cite{Lu2012,Zhang2021,Lu2024}. Together, these properties represent electronic stability, local electrostatic interactions and molecular polarity, but remain screening proxies rather than direct measures of condensed-phase electrolyte performance.

Even at the isolated-molecule level, evaluating these five descriptors by DFT for nearly one billion candidates would remain computationally prohibitive. Because the descriptors depend on molecular geometry and local atomic environments, their prediction requires a representation that resolves three-dimensional molecular structure. We selected Uni-Mol, a three-dimensional molecular model pretrained on large-scale conformer data to encode atomic environments, interatomic distances and molecular geometry\cite{UniMol2023,unimolv2}. An independent Uni-Mol model was fine-tuned for each DFT-level electronic--solvation descriptor.

Following the data-construction and DFT-labeling procedures detailed in the Methods section, we assembled a fine-tuning dataset of approximately $10^5$ DFT-labeled PubChem molecules after structure standardization, element filtering and restriction by heavy-atom count\cite{pubchem2021}. The dataset spans carbonyl, amine, halogenated, aromatic, ether, amide and hydroxyl groups, among other polar or reactive functionalities, with substantial fractions of O- and N-containing molecules, as summarized in Fig.~S1(a--c). Its heavy-atom-number and molecular-weight distributions also overlap those of the reference electrolyte set, as shown in Fig.~\ref{fig:dataset}(b) and Fig.~S2. To compare the structural spaces represented by the training set and GDB13, we encoded molecular substructures using Molecular ACCess System (MACCS) structural-key fingerprints\cite{durant2002} and projected both sets into two dimensions using t-distributed stochastic neighbor embedding (t-SNE)\cite{maaten2008}. The projected distributions overlap extensively, with most PubChem training molecules located within or adjacent to regions occupied by GDB13 molecules, as shown in Fig.~\ref{fig:dataset}(a). Because t-SNE preserves local neighborhoods rather than global distances, this projection provides only qualitative context for the structural relationship between the datasets. Transfer performance in the GDB13 candidate space is evaluated independently below.

Multidimensional scoring requires descriptors that provide complementary rather than redundant molecular information. As shown in Fig.~\ref{fig:dataset}(c), HOMO and LUMO have weak-to-moderate correlations with the ESP descriptors and MPI ($|r|\leq 0.37$), whereas MPI correlates positively with \ESPmax ($r=0.58$) and negatively with \ESPmin ($r=-0.51$). Frontier-orbital energies, local electrostatic extrema and molecular polarity thus constitute related but nonredundant coordinates. The joint distributions in Fig.~\ref{fig:dataset}(d--f) also show that the PubChem-derived DFT dataset spans broad ranges of all five target properties. Establishing these descriptor relationships and ranges, however, does not by itself demonstrate that they can be predicted accurately in GDB13.

We assessed model performance first within the PubChem-derived chemical space and then on GDB13 molecules. As shown in Fig.~\ref{fig:dataset}(g), predictions for the PubChem held-out test set closely follow the ideal diagonal, with $R^2$ values ranging from 0.9018 for \ESPmin to 0.9726 for HOMO. This evaluation measures performance within the chemical space used for model development rather than transfer to GDB13. Transfer was assessed using an independent test set composed of GDB13 molecules with no overlap with the PubChem fine-tuning dataset. As shown in Fig.~S3, the $R^2$ values are 0.9235 and 0.9084 for HOMO and LUMO, respectively, and 0.9816, 0.9812 and 0.9857 for \ESPmin, \ESPmax and MPI. The models thus retain high agreement with the DFT labels in the candidate space used for subsequent screening. These validated models provide the DFT-level proxy descriptors used to rank GDB13 candidates and construct the descriptor-weight motif-dominance map described in the following section.

\subsection{Descriptor-weight motif-dominance map reveals motif regimes}
\begin{figure}[htbp!]
  \centering
  \includegraphics[width=0.95\textwidth]{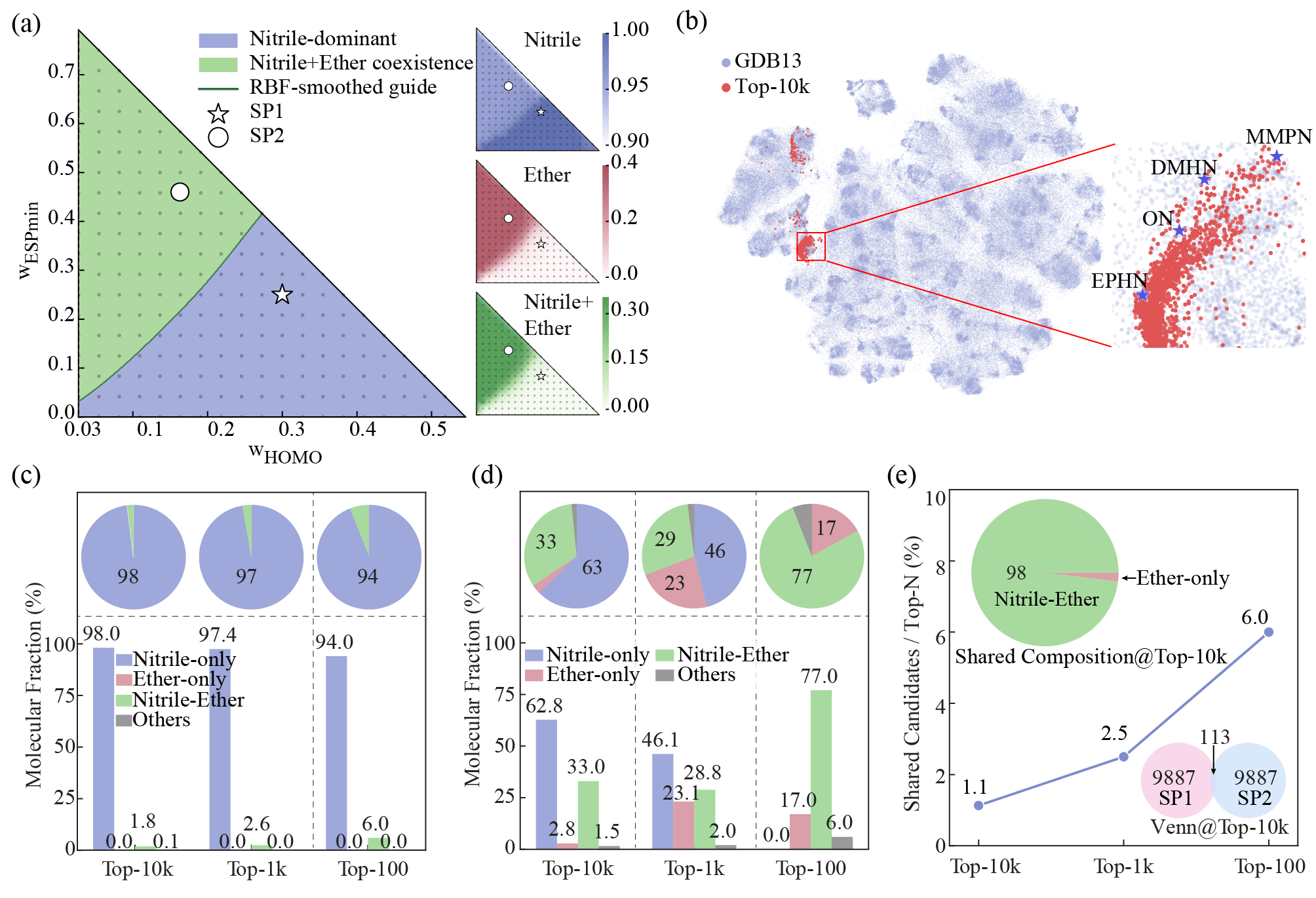}
  \caption{Descriptor-weight motif-dominance map and representative high-ranking molecular sets.
  (a) Motif-dominance map in the two-dimensional descriptor-weight space spanned by $w_{\mathrm{HOMO}}$ and $w_{\mathrm{ESP}_{\min}}$, based on non-exclusive nitrile-containing and ether-containing fractions in the Top-10k molecules. A nitrile-dominant point has a nitrile-containing fraction above 30\% and an ether-containing fraction no greater than 30\%; both fractions exceed 30\% in the nitrile + ether coexistence regime. The purely orbital-free edge is not included. The smaller maps show the corresponding Top-10k fractions of nitrile-containing, ether-containing, and nitrile--ether molecules. The solid curve is a radial-basis-function (RBF)-smoothed visual guide rather than a fitted physical boundary. SP1 and SP2 are representative scoring points in the nitrile-dominant and nitrile + ether coexistence regimes, respectively.
  (b) Two-dimensional chemical-space projection of GDB13 molecules and the SP1 Top-10k candidates, with a local enlargement containing representative nitrile- and nitrile--ether-containing structures.
  (c,d) Mutually exclusive Top-N molecular composition at SP1 and SP2, respectively, classified as nitrile-only, ether-only, nitrile--ether, and others.
  (e) Fraction of candidates shared between SP1 and SP2 as a function of the Top-N rank cutoff. The insets show the motif composition of the shared Top-10k candidates and a Venn diagram of the SP1 and SP2 Top-10k sets.}
  \label{fig:screening}
\end{figure}

To identify structural motifs favored under different combinations of electronic and solvation properties, we combined the five descriptors into a composite scoring function. The score contains no explicit reward for nitrile, ether, or any molecular scaffold and applies no functional-group filter. Changes in the high-ranking molecular populations can therefore be attributed to descriptor-level selection rather than a predefined structural preference.

For a given descriptor-weight combination, the composite score is
\begin{equation}
S = \sum_{d \in \mathcal{D}} w_d S_d,
\qquad \sum_{d \in \mathcal{D}} w_d = 1.
\label{eq:score}
\end{equation}
Here, $\mathcal{D}=\{\mathrm{HOMO}, \mathrm{LUMO}, \mathrm{ESP}_{\min}, \mathrm{ESP}_{\max}, \mathrm{MPI}\}$, while $S_d$ and $w_d$ denote the subscore and weight of descriptor $d$, respectively. Each subscore is bounded between 0 and 1. HOMO and LUMO use monotonic clipped functions because lower HOMO and higher LUMO values indicate progressively lower intrinsic oxidation and reduction tendencies until the reference range is reached\cite{Cheng2015,Pande2019,Gao2023,Peljo2018}. By contrast, \ESPmin, \ESPmax, and MPI use window functions because electrostatic and polarity descriptors distinguish different solvent-interaction regimes rather than defining a universally favorable monotonic direction\cite{Wu2023,Yao2021,Kim2021,Su2019}. The two subscore forms are
\begin{equation}
\begin{aligned}
S_{\mathrm{HOMO/LUMO}} &=
\mathrm{clip}\left(
\frac{\mp(\varepsilon-\varepsilon_0)}{\lambda_{\varepsilon}},0,1
\right),\\
S_{\mathrm{ESP/MPI}} &=
\max\left(
0,1-\frac{|p-p_0|}{\lambda_p}
\right).
\end{aligned}
\label{eq:score_terms_general}
\end{equation}
where $\mathrm{clip}$ bounds a value between 0 and 1. For HOMO and LUMO, $\varepsilon$ is the orbital energy, the upper and lower signs apply to HOMO and LUMO, respectively, $\varepsilon_0$ is the zero-score threshold, and $\lambda_{\varepsilon}$ is the transition width. For \ESPmin, \ESPmax, and MPI, $p_0$ is the window center and $\lambda_p$ is the window half-width.

We derived the subscore parameters in Table~\ref{tab:score_parameters} from a single reference set of 65 literature-reported electrolyte molecules using prespecified statistical rules independent of the high-ranking GDB13 candidates. After aligning the orbital-energy directions as $-E_{\mathrm{HOMO}}$ and $E_{\mathrm{LUMO}}$, the 20th and 80th percentiles of the reference distributions defined the zero- and full-score ends of each monotonic subscore. As shown in Fig.~S5(a,b), this procedure gave HOMO endpoints of $-7.0$ and $-8.2$ eV and LUMO endpoints of $-0.2$ and $1.0$ eV, corresponding to a transition width of $1.2$ eV in both cases. The full-score endpoints are close to the corresponding frontier-orbital energies of EC, aligning the percentile-derived anchors with the electronic-stability range of a representative carbonate electrolyte molecule. These transitions distinguish molecules outside the reference range without treating either endpoint as a strict physical stability boundary.

\begin{table}[htbp]
  \caption{Parameters of descriptor subscore functions derived from the reference electrolyte molecule set. The scale width denotes the transition width for monotonic subscores and the window half-width for window subscores.}
  \label{tab:score_parameters}
  \centering
  \setlength{\tabcolsep}{3pt}
  \begin{tabular}{lccc}
\toprule
\specialrule{0em}{0pt}{0.8pt}
\hline
Descriptor & Subscore Type & Threshold/Center & Scale Width \\
\midrule
HOMO & Monotonic & $-7.0$ eV & $1.2$ eV \\
LUMO & Monotonic & $-0.2$ eV & $1.2$ eV \\
\ESPmin & Window & $-35$ kcal/mol & $15$ kcal/mol \\
\ESPmax & Window & $18$ kcal/mol  & $15$ kcal/mol \\
MPI & Window & 12 & 12 \\
  \hline
  \specialrule{0em}{0pt}{0.8pt}
  \bottomrule
  \end{tabular}
\end{table}

For \ESPmin, \ESPmax, and MPI, the mean of each reference distribution defined the window center, and the 10th and 90th percentiles characterized its central range. The window half-width was twice the larger absolute deviation of these percentiles from the mean. As shown in Fig.~S5(c--e), this rule yielded centers of $-35$ kcal/mol, $18$ kcal/mol, and 12 and half-widths of 15 kcal/mol, 15 kcal/mol, and 12 for \ESPmin, \ESPmax, and MPI, respectively. Together, these prespecified parameters define a reference window rather than strict physical boundaries or motif-specific optima. Their finite widths retain resolution near the reference distributions without amplifying small descriptor variations, while preserving penalties for values far outside those ranges.

We varied the descriptor weights to determine whether the composition of high-ranking molecules persisted beyond a single score definition. Because local negative electrostatic potential is closely associated with electron-rich interaction sites\cite{Wu2023,Liu2018}, we retained stronger emphasis on \ESPmin within the ESP channel by imposing $w_{\mathrm{ESP}_{\min}}=5w_{\mathrm{ESP}_{\max}}$. Oxidation sensitivity and reduction- or interphase-related tendencies are both relevant to electrolyte selection\cite{Park2021,Zhang2006}; accordingly, the frontier-orbital weights were kept nearly balanced, with $w_{\mathrm{HOMO}}=1.2w_{\mathrm{LUMO}}$. Together with $\sum_{d\in\mathcal{D}}w_d=1$, these relations leave $w_{\mathrm{HOMO}}$ and $w_{\mathrm{ESP}_{\min}}$ as the two independent coordinates and assign the remaining weight to MPI.

At each retained weight point, lossless score-bound pruning produced a Top-10k ranking algorithmically equivalent to full-space ranking of GDB13. The purely orbital-free edge, $w_{\mathrm{HOMO}}=w_{\mathrm{LUMO}}=0$, was excluded because full-space equivalence could not be certified there from the score bounds used for pruning. Non-exclusive nitrile-containing and ether-containing fractions were calculated for each Top-10k set, with nitrile--ether molecules contributing to both fractions. A point was classified as nitrile-dominant when the nitrile-containing fraction exceeded 30\% and the ether-containing fraction did not; coexistence required both fractions to exceed 30\%.

The retained weight space separates into two connected regimes under this operational classification, as shown in Fig.~\ref{fig:screening}(a). Nitrile-containing molecules remain prevalent across the broad nitrile-dominant region, showing that their high-ranking status persists over a continuous range of electronic and electrostatic weightings. Increasing the relative emphasis on \ESPmin and MPI raises the ether-containing and nitrile--ether fractions and produces the nitrile + ether coexistence regime. Ether enrichment depends on the descriptor weighting and does not dominate independently across the map.

For the mutually exclusive composition analyses, molecules were classified as nitrile-only, ether-only, nitrile--ether, or others. The residual ``others'' class contains molecules with neither nitrile nor ether and is not a chemically homogeneous motif. Nitrile and ether were retained as the principal composition axes because their fractions show the clearest systematic changes across the weight map.

To determine whether other functional groups show comparable behavior across the descriptor space, Fig.~S6 compares their non-exclusive descriptor distributions in the score-evaluable GDB13 subset used to construct the map. Nitrile-containing molecules occupy ranges compatible with both frontier-orbital and electrostatic/polarity constraints, whereas ether and several other oxygen-containing groups show more pronounced differences in the ESP and MPI distributions. Other functional groups display favorable values in individual descriptors, but these ranges are generally more localized or descriptor-specific. This comparison provides context for the two principal regimes without defining additional motif classes.

We selected SP1 and SP2 as representative scoring points in the nitrile-dominant and nitrile + ether coexistence regimes, respectively. To locate the SP1 high-ranking molecules within the broader candidate space, we projected their MACCS fingerprints together with the GDB13 background using t-SNE. As shown in Fig.~\ref{fig:screening}(b), the SP1 Top-10k molecules occupy a localized region of the two-dimensional projection. In the enlarged view, these candidates lie near representative nitrile- and nitrile--ether-containing structures, including 2-ethyl-2-propylhexanenitrile (EPHN), octanenitrile (ON), 2,3-Dimethylhexanenitrile (DMHN), and 2-Methoxy-2-methylpropanenitrile (MMPN).

The mutually exclusive composition at SP1 quantifies the prevalence of nitrile-only molecules. As shown in Fig.~\ref{fig:screening}(c), they account for 98.0\%, 97.4\%, and 94.0\% of the Top-10k, Top-1k, and Top-100 sets, respectively. Fig.~S7 gives a GDB13 background fraction of only 4.7\%, while Fig.~S8(a) shows enrichment factors above 20 at all three rank cutoffs. Ether-only molecules are nearly absent, while the nitrile--ether fraction increases modestly from 1.8\% in the Top-10k set to 6.0\% in the Top-100 set. Nitrile-only molecules remain the principal high-ranking class at SP1 even as the rank cutoff is tightened.

The corresponding composition at SP2 shows a sustained decline in nitrile-only molecules and an eventual dominance of the nitrile--ether class at the strictest rank cutoff. Fig.~\ref{fig:screening}(d) shows that the Top-10k set contains 62.8\% nitrile-only and 33.0\% nitrile--ether molecules. At Top-1k, these fractions become 46.1\% and 28.8\%, respectively, while ether-only molecules account for 23.1\%. In the Top-100 set, nitrile-only molecules are absent, whereas nitrile--ether and ether-only molecules reach 77.0\% and 17.0\%, respectively. Relative to their GDB13 background, the enrichment factor of nitrile--ether molecules increases from 11.8 at Top-10k to 27.5 at Top-100, as shown in Fig.~S8(b).

The descriptor and subscore distributions provide the basis for interpreting this composition shift. As shown in Fig.~S9, the four mutually exclusive classes occupy different regions of the five-descriptor space within the score-evaluable GDB13 subset. At SP1, the high-ranking molecules retain strong frontier-orbital subscores together with a moderate \ESPmin contribution, whereas those at SP2 approach saturation in the ESP- and MPI-related subscores, as shown in Fig.~S10. This redistribution of descriptor-level selection pressure accounts for the transition between regimes. Nitrile-containing molecules remain competitive across these conditions, while ether contributes conditionally by tuning polarity- and solvation-related descriptors as their weights increase.

Comparing the candidates shared by SP1 and SP2 identifies the molecular classes that remain competitive under both descriptor-weight conditions. For each rank cutoff, the overlap fraction was defined as the number of shared molecules divided by $N$, where $N$ is the number of molecules in the corresponding Top-N set. As shown in Fig.~\ref{fig:screening}(e), the overlap fractions are 1.1\%, 2.5\%, and 6.0\% for Top-10k, Top-1k, and Top-100, respectively, showing that the two scoring points select largely distinct molecular populations. Among the shared Top-10k candidates, 98\% are nitrile--ether molecules, with only a minor ether-only fraction. The shared Top-1k and Top-100 sets consist entirely of nitrile--ether molecules. This concentration of shared candidates in the nitrile--ether class establishes its role as a structural bridge between the two regimes and supports the nitrile-centered, ether-assisted motif hierarchy.

\subsection{Interpretability analysis links motif enrichment to descriptor-level rules}

\begin{figure}[htbp!]
  \centering
  \includegraphics[width=0.95\textwidth]{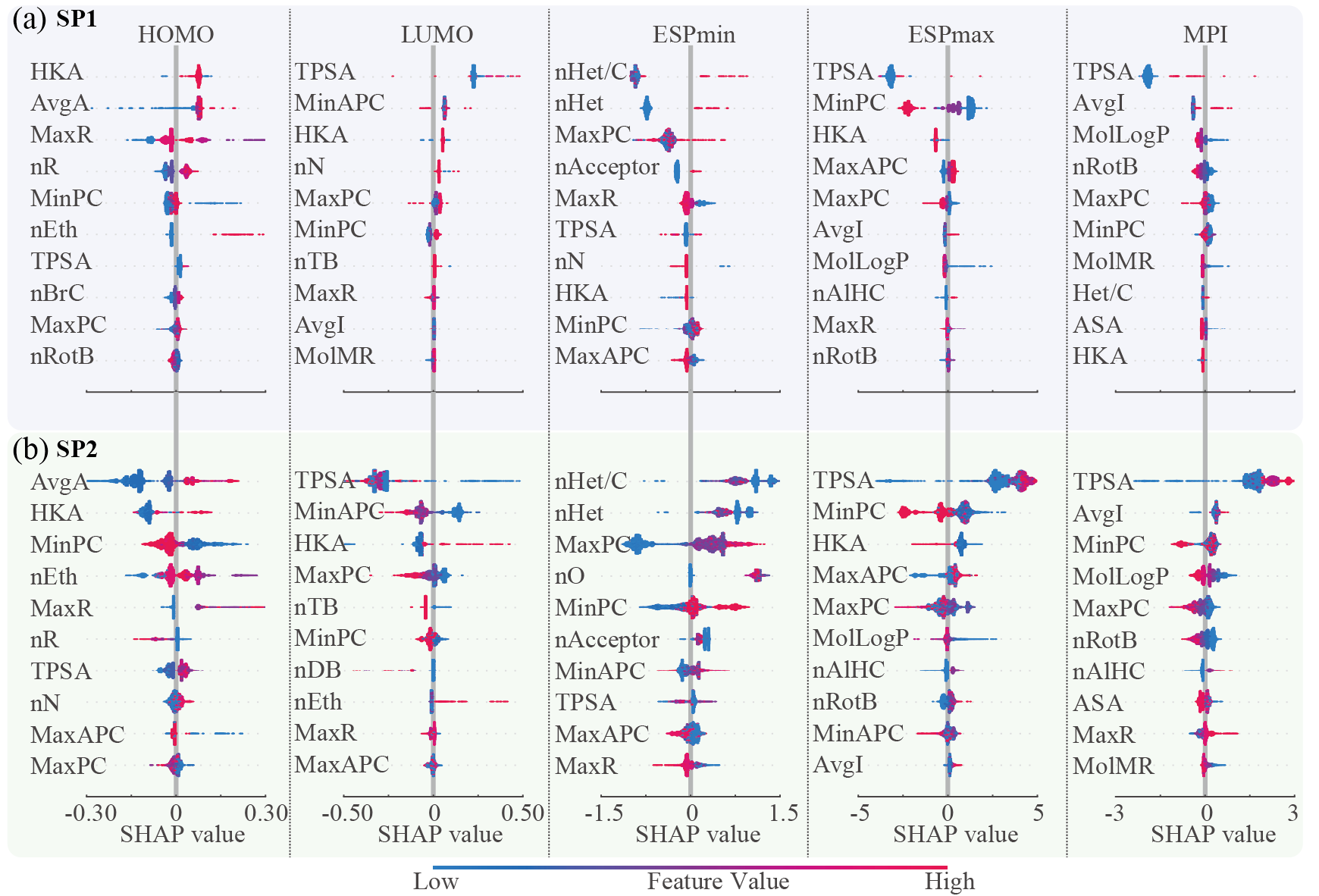}
  \caption{SHAP analysis of descriptor-level structural controls at representative scoring points.
  (a,b) SHAP beeswarm plots for the five DFT-level proxy descriptors at SP1 and SP2, respectively. SP1 is selected from the nitrile-dominant region, whereas SP2 is selected from the nitrile + ether coexistence region in the descriptor-weight motif-dominance map. For each target descriptor, structural features are ranked by SHAP importance, point color denotes the corresponding feature value, and each point represents one molecule.}
  \label{fig:shap}
\end{figure}

The descriptor-weight map establishes a nitrile-centered, ether-assisted motif hierarchy at the population level, but its structural basis must be resolved through the descriptors that define the scoring window. We therefore trained one XGBoost interpretation model\cite{chen2016xgboost} for each proxy descriptor using the combined SP1 and SP2 Top-10k molecules and evaluated SHapley additive explanations (SHAP)~\cite{lundberg2020} separately for the two high-ranking sets. This common modeling basis makes the SHAP values for a given proxy descriptor comparable between SP1 and SP2. The input features span atomic composition, bonding and topology, functional groups, local partial charges and electronic properties, as defined in Table~S2. The feature-correlation matrix shown in Fig.~S11 identifies chemically related local modules, while preserving distinct structural dependencies for the five proxy descriptors.

The SHAP distributions in Fig.~\ref{fig:shap}(a,b) separate the structural controls of frontier-orbital energies from those of surface electrostatics and polarity. HOMO is governed primarily by electronic features, local partial-charge extrema and molecular size or topology, whereas LUMO is more sensitive to polar surface area, heteroatom composition, multiple bonds and local charge. Neither orbital energy is controlled by a simple nitrile count. Instead, the electronic-stability contribution arises from the combined effects of electron-withdrawing fragments, heteroatom composition, charge distribution and molecular topology.

The solvation-related descriptors depend more directly on distributed heteroatom and surface properties. \ESPmin is associated mainly with heteroatom density, acceptor-site abundance, polar surface area and local partial charges. Oxygen count becomes more influential for \ESPmin among the high-ranking molecules at SP2, indicating a stronger contribution from O-containing polar structures without, by itself, assigning that contribution exclusively to ether groups. \ESPmax responds to polar surface area, charge separation and surface character, while MPI reflects polar surface area, lipophilicity, molecular size, flexibility and partial-charge extrema. The complete distributions in Figs.~S12 and S13 show that substantial SHAP contributions remain concentrated in these chemically meaningful feature families rather than being dispersed over numerous weakly relevant variables.

The quantitative summaries in Fig.~S14 connect these descriptor-level controls to the composition shift between SP1 and SP2. Electronic, N-containing, local-charge and topological features support the nitrile-only high-score core at SP1. As the ESP/MPI-related constraints become stronger at SP2, polar surface area, oxygen content, ether count and local-charge features contribute more strongly across several proxy descriptors. The motif-level descriptor \(z\)-scores show the same complementarity: nitrile-containing classes retain nitrile and N-related features, whereas ether-containing and nitrile--ether classes exhibit greater O/C ratios, oxygen and ether counts, heteroatom density and polar surface area. This structural evidence is consistent with the composition change in Fig.~\ref{fig:screening}, from a nitrile-only-dominated population at SP1 to increasing nitrile--ether prevalence among the highest-ranking molecules at SP2.

The shared nitrile--ether candidates consequently preserve the nitrile-associated electronic and N-related electrostatic contributions while adding O-related degrees of freedom for tuning polarity and solvation. The motif hierarchy is therefore not imposed during screening through functional-group rewards; it arises from the relationship among molecular structure, the DFT-level proxy descriptors and the multi-objective scoring window. This result defines both an electronic--solvation objective and a corresponding structural hierarchy, providing the basis for testing whether the extracted rule can guide molecular generation beyond the ranked GDB13 library.

\subsection{Rule-guided generative optimization expands the candidate space}
\begin{figure}[htbp!]
  \centering
  \includegraphics[width=0.95\textwidth]{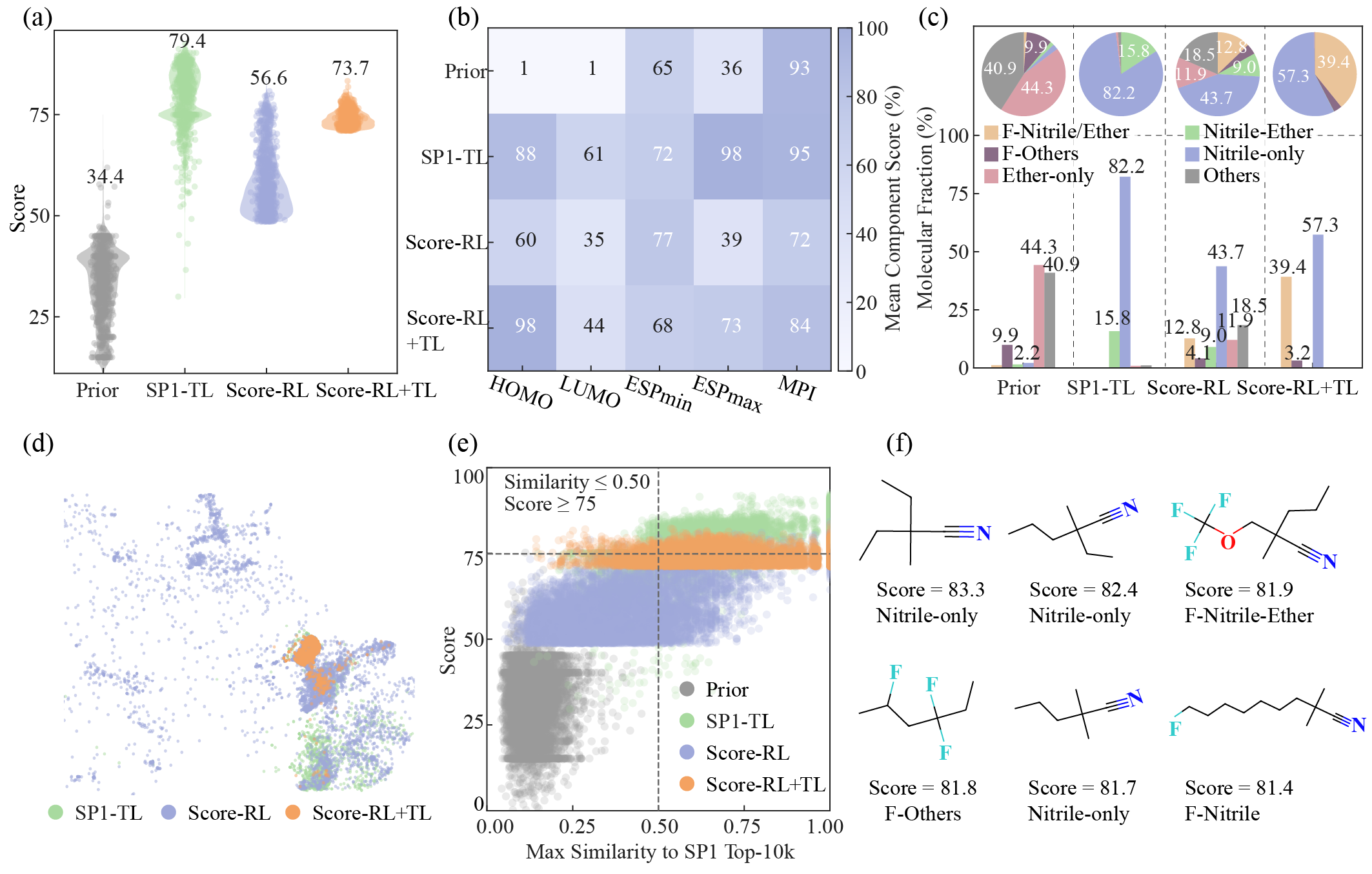}
  \caption{Generative-model transfer and rule-guided extrapolation.
  (a) Score distributions of molecules sampled from the original REINVENT prior, the SP1 Top-10k transfer-learning model (SP1-TL), the rule-guided reinforcement-learning model combining the electronic--solvation score with a nitrile anchor (Score-RL), and the post-RL transfer-learning model trained on high-scoring Score-RL molecules (Score-RL+TL). Score-RL was initialized from an electrolyte-domain-adapted prior trained on 65 literature electrolyte molecules.
  (b) Mean component scores for the five descriptor subscore terms in Prior, SP1-TL, Score-RL and Score-RL+TL samples. 
  (c) Motif composition of generated molecules. F-Nitrile/Ether is a pooled category comprising F-containing molecules with nitrile motifs, ether motifs, or both, whereas F-Others comprises F-containing molecules with neither motif.
  (d) Chemical-space projection of generated molecules from SP1-TL, Score-RL and Score-RL+TL. 
  (e) Relationship between total score and maximum molecular similarity to SP1 Top-10k. Dashed lines indicate score 75 and similarity 0.50. 
  (f) Representative high-scoring generated molecules with structure-specific motif assignments.
  }
  \label{fig:rl}
\end{figure}

The motif-dominance map and SHAP analysis provide both the electronic--solvation objective and the nitrile-centered, ether-assisted hierarchy used for molecular generation. Two routes were compared. Transfer learning on the SP1 Top-10k molecules (SP1-TL) transfers the complete high-ranking GDB13 distribution, including its descriptor profile and structural composition. Score-RL instead implements the extracted rule through a composite reward that combines the electronic--solvation score with a nitrile anchor. Because nitrile-containing structures are sparse in the original, non-electrolyte-specific REINVENT prior\cite{loeffler2024reinvent}, the prior was first adapted using the 65 literature-reported electrolyte molecules, without exposure to the SP1 Top-10k set. The nitrile anchor thereby supplies the structural component of the rule, rather than serving as an independent test of nitrile enrichment.

The total- and component-score distributions distinguish direct distribution transfer from reward-guided optimization. As shown in Fig.~\ref{fig:rl}(a), the mean score increases from 34.4 for the original prior to 79.4 after SP1-TL. Score-RL raises the mean to 56.6, and subsequent transfer learning on Score-RL molecules with scores of at least 70 increases it to 73.7. The component scores in Fig.~\ref{fig:rl}(b) show that these routes reach different regions of the objective. SP1-TL improves all five subscores by learning directly from SP1 candidates. Score-RL improves the \ESPmin and MPI subscores relative to the prior but remains lower than SP1-TL in LUMO and \ESPmax. Score-RL+TL consolidates the high-scoring RL population by increasing its HOMO, \ESPmax and MPI subscores while retaining a moderate \ESPmin contribution. Thus, reward-guided optimization and post-RL transfer learning produce a descriptor compromise distinct from direct reproduction of the SP1 distribution.

The corresponding motif compositions reveal how the structural hierarchy is realized in each generated population. In Fig.~\ref{fig:rl}(c), SP1-TL produces 82.2\% nitrile-only and 15.8\% nitrile--ether molecules, consistent with direct transfer from the F-free SP1 set. Score-RL yields 43.7\% nitrile-only molecules together with ether-only, nitrile--ether and F-containing structures; its F-Nitrile/Ether and F-Others fractions are 12.8\% and 4.1\%, respectively. Score-RL+TL increases the nitrile-only and F-Nitrile/Ether fractions to 57.3\% and 39.4\%, while the F-Others fraction remains 3.2\%. Nitrile enrichment follows directly from the explicit anchor, whereas fluorination is not encoded in the reward. The preferential retention of F among nitrile- and/or ether-containing molecules is therefore an unprescribed response of the generator to the descriptor objective. Because GDB13 contains no F atoms, these structures also test whether the extracted descriptor rule can be applied beyond the element space from which the motif hierarchy was inferred.

This extrapolation is chemically plausible but does not establish fluorination as a generally favorable modification. Fluorination can alter frontier-orbital energies, \Li coordination, solvation structure and LiF-containing interphase chemistry~\cite{Li2023,Fan2021,Fan2018,Yu12020,Jin2017,Michan2016}, but may also weaken salt solvation, reduce ionic conductivity, increase viscosity or promote phase and interphase instabilities~\cite{Li2023,Li2025}. The F-containing molecules are consequently prospective outcomes of the rule-guided search, not additional evidence for the nitrile--ether hierarchy identified in GDB13.

Chemical-space and similarity analyses further separate generative expansion from retrieval of ranked candidates. The projection in Fig.~\ref{fig:rl}(d) shows broader coverage by Score-RL than by SP1-TL, while Score-RL+TL concentrates within the region accessed during RL. Consistently, Fig.~\ref{fig:rl}(e) contains Score-RL and Score-RL+TL molecules with scores above 75 and maximum similarities below 0.50 to the SP1 Top-10k set. The representative structures in Fig.~\ref{fig:rl}(f) retain the anchored nitrile motif across varied sizes, branching patterns and topologies, and include fluorinated nitrile and ether variants outside the GDB13 element space. Rule-guided generation therefore extends the hierarchy from population-level analysis to prospective molecular exploration. The generated FN candidate provides one test of this extension in the condensed phase, alongside literature benchmarks and representative high-ranking GDB13 molecules.

\subsection{Condensed-phase behavior and boundaries of weak coordination}

The preceding analyses identified a nitrile-centered, ether-assisted motif hierarchy from physically motivated electronic--solvation descriptors. Its condensed-phase implications were examined by molecular dynamics (MD) simulations at additive concentrations of 5\%, 10\% and 20\% in an ethylene carbonate (EC) electrolyte containing 1.0 M Li\PF. Valeronitrile (VN) and ON served as literature-reported nitrile benchmarks\cite{wang2024vn,gao2026}. From the SP1 high-ranking set, EPHN and 1,2:4,5-diepoxypentane (DEP) represented nitrile-only and ether-only structures, respectively. MMPN, shared by the SP1 and SP2 Top-10k sets, represented the nitrile--ether bridge class. The generated fluorinated nitrile, 2-(fluoromethyl)-2-methylheptanenitrile (FN), provided a test beyond the GDB13 element space. The simulations resolved whether these additives entered the local \Li environment while preserving EC in the first solvation shell, and how their participation affected \Li--\PF association and ion transport.

\begin{figure}[htbp!]
  \centering
  \includegraphics[width=0.95\textwidth]{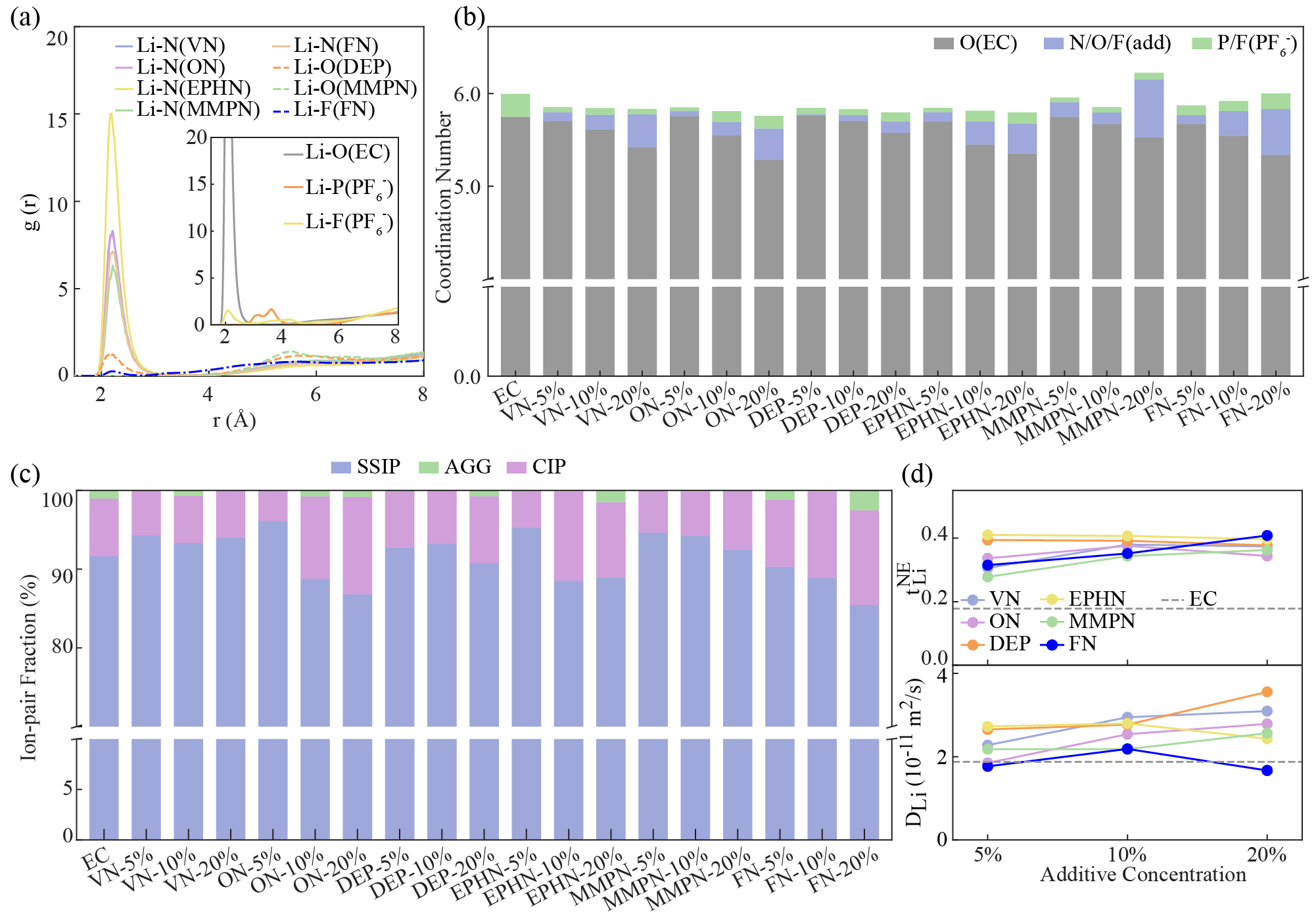}
  \caption{Condensed-phase MD analysis of representative additive-containing electrolytes.
  (a) Radial distribution functions between \Li and the N, O, and F sites of the six additives at 10\% additive concentration; the inset shows reference RDFs for \Li--O pairs involving EC and \Li--P/\Li--F pairs involving \PF.
  (b) Coordination-number contributions from O atoms in EC, N, O, and F sites in the additives, and P and F atoms in \PF at different additive concentrations.
  (c) Fractions of solvent-separated ion pairs (SSIP), contact ion pairs (CIP), and aggregates (AGG) in the same simulated systems.
  (d) Nernst--Einstein \Li transference numbers, \(t_{\mathrm{Li}}^{\mathrm{NE}}\), and \Li diffusion coefficients, \(D_{\mathrm{Li}}\), for different additives and additive concentrations; dashed lines denote the EC references.}
  \label{fig:md}
\end{figure}

Local solvation was evaluated by combining radial distribution functions (RDFs) with first-shell coordination numbers. The RDFs in Fig.~\ref{fig:md}(a) identify the additive sites that approach \Li and their characteristic distances, whereas the decomposition in Fig.~\ref{fig:md}(b) quantifies their contributions to the first solvation shell. At 10\% additive concentration, all nitrile-containing molecules, including FN, produced a \Li--N peak near 2.2~\AA{}, while the \Li--O RDF of the ether-only DEP showed a substantially weaker first-shell feature. The strongest first-shell peak nevertheless remained the \Li--O correlation associated with EC, and the complete concentration series in Fig.~S15 showed the same pattern. Consistent with this RDF picture, the total coordination number remained close to six, with O atoms in EC providing the largest contribution. Additive-site contributions generally increased with concentration, most notably for MMPN and FN, but remained a minority even at 20\%. The representative additives therefore entered the local \Li environment without replacing the EC-dominated first solvation shell.

This limited first-shell participation nevertheless altered \Li--\PF association in a molecule- and concentration-dependent manner. The solvent-separated ion-pair (SSIP), contact-ion-pair (CIP) and aggregate (AGG) fractions in Fig.~\ref{fig:md}(c) show that SSIPs remained dominant and AGG fractions remained small across all simulated systems. The CIP fraction, however, did not follow a common concentration trend. VN maintained a high SSIP fraction throughout the investigated range, whereas ON showed a marked increase in CIP fraction at higher concentrations. EPHN and FN also developed larger CIP/AGG fractions at higher concentrations, with 20\% FN deviating most strongly from the EC reference. Retention of an EC-dominated first shell therefore did not ensure progressively weaker \Li--\PF contact as the additive concentration increased. The effect depended on both molecular skeleton and concentration.

The resulting changes in ion association did not translate directly into \Li mobility. Fig.~\ref{fig:md}(d) compares the Nernst--Einstein (NE) transference number with the \Li self-diffusion coefficient. Although all additive-containing systems showed transference numbers above the EC reference, their \Li diffusion coefficients followed different concentration trends. VN and DEP exhibited higher \(D_{\mathrm{Li}}\) over most or all of the investigated range, whereas EPHN and MMPN showed non-monotonic responses. FN provided the clearest separation between the two transport measures: its \(t_{\mathrm{Li}}^{\mathrm{NE}}\) increased with concentration, while \(D_{\mathrm{Li}}\) fell below the EC reference at 20\%. Because \(t_{\mathrm{Li}}^{\mathrm{NE}}=D_{\mathrm{Li}}/(D_{\mathrm{Li}}+D_{\mathrm{PF}_6})\), the accompanying decrease in \(D_{\mathrm{PF}_6}\) shown in Fig.~S16 indicates that the high FN transference number arose partly from slower anion diffusion rather than solely from enhanced \Li mobility. The comparison with additive-site coordination in Fig.~S17 further shows that systems with similar coordination numbers can exhibit different transport responses. Local additive participation is therefore not a sufficient predictor of ion transport; molecular structure, ion association and the relative mobilities of the cation and anion remain coupled.

Within the simulated EC-based systems and investigated concentration range, MMPN and FN provide complementary tests of the motif hierarchy. MMPN, selected from candidates shared by SP1 and SP2, exhibited limited additive-site coordination while preserving the EC-dominated first shell. FN, generated outside the GDB13 element space, showed a comparable coordination pattern despite its F substitution. Their ion-association and transport responses, however, remained dependent on molecular identity and concentration. The MD results therefore support a bounded weak-coordination principle in which local solvation, ion association and transport remain jointly dependent on molecular structure and concentration.

\section{Conclusion}

This work develops a route for extracting electrolyte-molecule design rules from billion-scale chemical space rather than treating high-throughput screening solely as candidate ranking. Five electronic--solvation descriptors were combined with Uni-Mol inference to construct a descriptor-weight landscape across GDB13 without assigning explicit rewards to nitrile, ether or any predefined molecular skeleton. The resulting landscape revealed a nitrile-centered, ether-assisted motif hierarchy. Nitrile motifs remained enriched across broad regions of descriptor weight space, whereas ether motifs became co-enriched with nitrile under stronger electrostatic and polarity constraints. Candidate-overlap and SHAP analyses further connected these regimes to coupled structural controls on frontier-orbital stability, local electrostatic potential and molecular polarity.

The extracted rule could also serve as an objective for molecular generation. SP1-TL transferred the high-ranking GDB13 distribution into a generative model, while rule-guided reinforcement learning combined the electronic--solvation score with a nitrile anchor and generated candidates beyond the original GDB13 element space, including fluorinated nitrile-containing structures. Explicit-solvation MD simulations then showed that representative nitrile-only and nitrile--ether molecules could participate in the local \Li environment without displacing EC from the first solvation shell. Their effects on \Li--\PF association and ion transport nevertheless varied with molecular skeleton and additive concentration. Together, these simulations support a bounded weak-coordination principle rather than an unconditional reward for a particular functional group.

The principal outcome is thus a framework that converts established knowledge of nitrile and ether chemistry into a quantitative design rule with an explicit descriptor basis and identifiable applicability boundaries. It links large-scale molecular ranking, interpretable structural analysis, generative exploration and condensed-phase evaluation while preserving the distinction between proxy-level selection and electrolyte-level behavior. Experimental measurements of electrochemical stability, ionic conductivity, interphase formation and cell performance will be required to determine how the descriptor-level rule and simulated condensed-phase trends translate into electrolyte and cell behavior under operating conditions.

\section*{Methods}
\noindent\textbf{Molecular dataset.} PubChem molecules\cite{pubchem2021} were standardized with RDKit 2025.9.3\cite{rdkit2025} by validating molecular graphs, converting valid structures to canonical simplified molecular-input line-entry system (SMILES) strings and removing duplicates. The dataset was restricted to neutral organic molecules containing C, H, N, O, S or halogen elements and no more than 20 heavy atoms. Structures that could not be parsed or assigned chemically valid valences were discarded. This procedure produced 99,644 molecules with successful DFT labels, of which 79,715 were assigned to model development and the remainder to held-out evaluation. GDB13\cite{blum2009}, which contains enumerated organic molecules with no more than 13 heavy atoms, was used as the large candidate space. A separate set of 65 literature-reported electrolyte molecules served as the reference for deriving the descriptor subscore parameters and adapting the starting model for rule-guided generation.

\noindent\textbf{Quantum-chemical calculations.} Three-dimensional coordinates from the PubChem structure-data file (SDF) records were used to initialize gas-phase calculations for neutral singlet molecules. Geometry optimization and frequency calculations were performed with Gaussian 16 Revision C.02\cite{gaussian16} using the B3LYP functional\cite{becke1993,lee1988}, the 6-311G(d,p) basis set\cite{krishnan1980}, and D3 dispersion with Becke--Johnson damping\cite{grimme2010,grimme2011}. HOMO and LUMO energies were extracted from the optimized calculations. The corresponding formatted checkpoint files were analyzed with Multiwfn 3.8(dev)~\cite{Lu2012,Lu2024} to obtain the minimum and maximum molecular-surface electrostatic potentials, \ESPmin and \ESPmax, and the molecular polarity index, MPI. These five quantities constituted the DFT-level electronic--solvation descriptor set.

\noindent\textbf{Uni-Mol training.}
An independent Uni-Mol v2 regression model was fine-tuned for each descriptor~\cite{UniMol2023,unimolv2}. The model-development set was subjected to the default random five-fold training procedure implemented in Uni-Mol Tools, while the preassigned PubChem test molecules were excluded from model development and used for independent evaluation. The HOMO, LUMO, \ESPmin, \ESPmax and MPI datasets each contained 79,715 model-development molecules and 19,929 test molecules. Each model was fine-tuned for 100 epochs with a batch size of 16 and a learning rate of $1\times10^{-4}$. An additional DFT-labeled GDB13 test set was used to assess transfer from the PubChem-derived data to the enumerated candidate space. The five-fold models were then used as ensembles for GDB13 inference. To reduce the computational cost of billion-scale prediction, molecules whose score upper bounds precluded entry into the Top-10k set were discarded before all five descriptor predictions were completed.

\noindent\textbf{Scoring, motif classification and interpretation.}
The predicted descriptors were converted to normalized subscores using the monotonic and window functions defined in Eqs.~\ref{eq:score} and \ref{eq:score_terms_general}. Subscore thresholds, centers and scale widths were derived from fixed quantiles or means of the 65-molecule reference electrolyte set, as summarized in Table~\ref{tab:score_parameters}. The descriptor-weight map was constructed under $w_{\mathrm{HOMO}}=1.2w_{\mathrm{LUMO}}$ and $w_{\mathrm{ESP}_{\min}}=5w_{\mathrm{ESP}_{\max}}$, with $w_{\mathrm{HOMO}}$ and $w_{\mathrm{ESP}_{\min}}$ spanning the two independent axes and $w_{\mathrm{MPI}}$ taking the remaining normalized weight. At each retained weight point, the Top-10k molecules were obtained by lossless score-bound pruning that is algorithmically equivalent to ranking the full GDB13 space. The purely orbital-free edge, $w_{\mathrm{HOMO}}=w_{\mathrm{LUMO}}=0$, was omitted because this equivalence could not be certified from the available score bounds.
Functional groups were identified by SMILES arbitrary target specification (SMARTS) matching with RDKit\cite{rdkit2025}. The motif-dominance map used non-exclusive nitrile-containing and ether-containing fractions, such that a molecule containing both motifs contributed to both fractions. A motif was designated dominant when its Top-10k fraction exceeded 30\%; simultaneous exceedance defined a coexistence region. Separate composition analyses used four mutually exclusive classes: nitrile-only, ether-only, nitrile--ether and others. Enrichment factors were calculated relative to the corresponding motif fractions in the GDB13 background. MACCS fingerprints~\cite{durant2002} were used for molecular similarity calculations and t-SNE projections~\cite{maaten2008}. For structural interpretation, XGBoost models\cite{chen2016xgboost} were trained separately for the five proxy descriptors on the combined SP1 and SP2 Top-10k sets. SHAP values~\cite{lundberg2020} for the two sets were subsequently evaluated against the same descriptor-specific models, allowing their structural contributions to be compared on a common basis.

\noindent\textbf{Rule-guided molecular generation.}
Molecular generation was performed with REINVENT4~\cite{loeffler2024reinvent} using three routes in addition to sampling from the original prior. SP1-TL directly fine-tuned the prior on the SP1 Top-10k molecules and therefore transferred the complete high-ranking GDB13 distribution. Score-RL was constructed independently of SP1-TL. Because the original REINVENT prior was not electrolyte-specific, it was first adapted for 30 epochs using the 65-molecule reference electrolyte set. Reinforcement learning was then performed with the staged-learning workflow and the difference between augmented and posterior (DAP) strategy. The reward was the geometric mean of the electronic--solvation score and a nitrile anchor, thereby encoding both the descriptor objective and the core motif identified from GDB13. The respective reward weights were 1 and 5 during the first 100 steps and 1 and 1 during the subsequent 300 steps. The DAP parameter $\sigma$ was 32, the learning rate was $2\times10^{-5}$ and the batch size was 64; unique-sequence filtering and randomized SMILES were enabled. Score-RL+TL was obtained by further fine-tuning the resulting RL model for three epochs on generated molecules with scores of at least 70. Generated molecules were compared by total and component scores, motif composition, chemical-space distribution and maximum fingerprint similarity to the SP1 Top-10k set.

\noindent\textbf{Molecular dynamics simulations.}
MD simulations were performed with GROMACS 2025.1\cite{abraham2015gromacs} for VN, ON, EPHN, DEP, MMPN and the generated FN candidate in 1.0 M Li\PF/EC. Additive molecular fractions of 5\%, 10\% and 20\% were defined relative to the total number of EC and additive molecules. Initial periodic configurations were assembled with PACKMOL 21.0.4\cite{martinez2009packmol}. Bonded and Lennard--Jones parameters for EC and the additives followed the optimized potentials for liquid simulations all-atom (OPLS-AA) force field\cite{jorgensen1996opls}, with atomic partial charges obtained by restrained electrostatic potential (RESP) fitting to quantum-chemical electrostatic potentials\cite{bayly1993resp}. The \Li\cite{jensen2006} and \PF\cite{lopes2004} parameters were adopted from published electrolyte force fields. A time step of 1 fs was used throughout. Electrostatic interactions were treated with particle-mesh Ewald summation\cite{essmann1995pme}, and real-space electrostatic and van der Waals cutoffs were both set to 1.5 nm.
Each system was energy-minimized and equilibrated at 298 K by 0.2 ns in the canonical (NVT) ensemble, followed by 1 ns of Berendsen-coupled\cite{berendsen1984} and 1 ns of Parrinello--Rahman\cite{parrinello1981} dynamics in the isothermal--isobaric (NPT) ensemble at 1 bar. The systems were subsequently annealed under NVT conditions from 298 to 400 K and back to 298 K over 3 ns, followed by a further 1 ns NPT equilibration at 298 K. Production trajectories were generated for 5 ns in the NVT ensemble at 298 K using a Nos\'e--Hoover thermostat\cite{nose1984,hoover1985}. Radial distribution functions and first-shell coordination numbers were calculated for O atoms in EC, N/O/F sites in the additives and P/F sites in \PF. SSIP, CIP and AGG populations were classified according to the number of \PF anions within a 5.0~\AA{} \Li--P cutoff of each \Li. Self-diffusion coefficients were obtained from linear fits to the central 30--80\% interval of the mean-square-displacement trajectories, and Nernst--Einstein transference numbers were calculated from the \Li and \PF diffusion coefficients.

\section*{Data Availability Statement}
The GDB13 and PubChem source data are publicly available. The processed datasets and derived data supporting this study, including the ranked candidate lists, trained model parameters and generated molecular structures, are available from the corresponding author upon reasonable request.

\section*{Supporting Information}
Composition and external validation of the molecular datasets; reference electrolyte set and descriptor-subscore parameters; supplementary motif and structural-interpretation analyses; additional molecular-generation results; molecular-dynamics methodology and extended solvation, ion-association, and transport results (PDF).

\section*{Author Information}
\subsection*{Corresponding Authors}
Zheng Cheng: chengz@aisi.ac.cn\\
Jinzhe Zeng: jinzhe.zeng@ustc.edu.cn\\
Qiangqiang Gu: guqq@ustc.edu.cn

\subsection*{Author Contributions}
Q.G. conceived and led the project. J.Z. and Z.C. co-supervised the research. Y.X. performed all calculations and model training. Y.X. and Q.G. wrote the manuscript, and all authors commented on the manuscript. All authors have read and approved the final manuscript.

\section*{Acknowledgment}
This work was supported by the National Natural Science Foundation of China (12504285), the National Science and Technology Advanced Materials Major Program of China (2025ZD0618401), and the Natural Science Foundation of Jiangsu Province (BK20250472). The numerical calculations in this paper have been done on the supercomputing system in the Supercomputing Center of University of Science and Technology of China.

\printbibliography[title={References}]

\clearpage
\section*{Table of Contents Graphic}
\begin{center}
  \includegraphics[width=3.25in,height=1.75in,keepaspectratio]{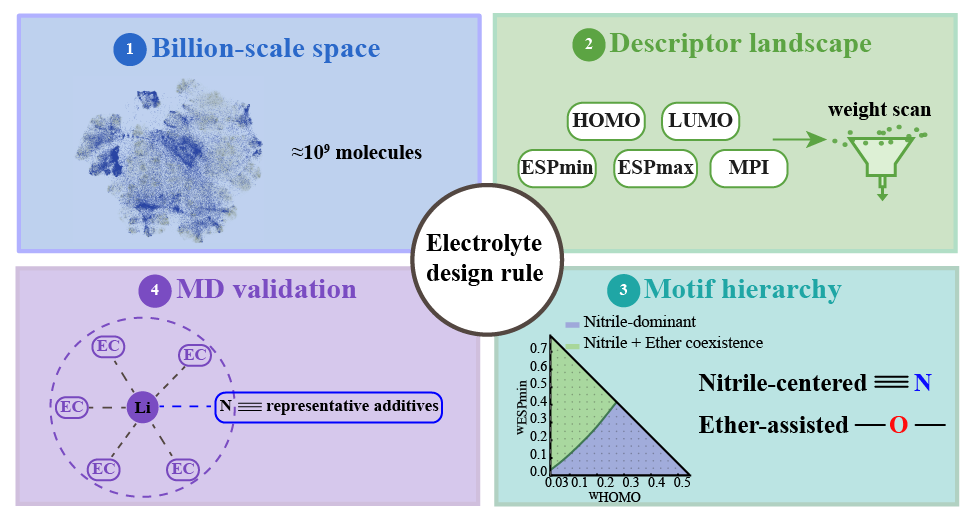}
\end{center}

\end{document}